\documentclass[10pt,journal,compsoc]{IEEEtran}

\usepackage{framed,multirow} % ??

\usepackage{amssymb}
\usepackage{latexsym}
\usepackage[latin9]{inputenc}
\usepackage{amsmath}
\usepackage{graphicx}
\usepackage{epstopdf} % converting to PDF
\usepackage{amssymb}  % pour mathbb
\usepackage{xcolor}   % for the colors (if necessary)
\usepackage{hyperref} % Pour href

\usepackage{svg}                       % Mehdi
\usepackage[ruled,vlined,linesnumbered]{algorithm2e} % Mehdi http://ctan.org/pkg/algorithm
\usepackage{algpseudocode}             % Mehdi http://ctan.org/pkg/algorithmicx
\usepackage{pdfpages}                  % Mehdi
\usepackage{url}
\usepackage{xcolor}

\usepackage{hyperref}

\definecolor{newcolor}{rgb}{.8,.349,.1}

\begin{document}
%\begin{frontmatter}
%\verso{Mehdi Yedroudj \textit{et~al.}}

\title{
{\footnotesize \textit{Accepted to Journal of Visual Communication and Image Representation, Elsevier, 2019 - version of september 2020.\vspace{+1.2cm}\\}}
Steganography using a 3-player game}%

%\author[1]{Mehdi \snm{Yedroudj}\fnref{fn1}}
%\fntext[fn1]{Mehdi.Yedroudj@lirmm.fr}
  %
%\author[1]{Fr\'ed\'eric \snm{Comby}\fnref{fn2}}
%\fntext[fn2]{Frederic.Comby@lirmm.fr}
%
%\author[1,2]{Marc  \snm{Chaumont}\fnref{fn3}}
%\fntext[fn3]{ Marc.Chaumont@lirmm.fr}
%
%\address[1]{LIRMM, CNRS, Univ. Montpellier, 161 rue Ada, Montpellier, 34095, France,}
%\address[2]{Univ. N\^{i}mes, 7 Place Gabriel P\'eri, 30000, France,}

\author{\IEEEauthorblockN{Mehdi YEDROUDJ}\\
{\scriptsize \IEEEauthorblockA{LIRMM, Univ. Montpellier, CNRS, Montpellier, France, Mehdi.Yedroudj@lirmm.fr}}\\
\and
\IEEEauthorblockN{Fr\'ed\'eric COMBY}\\
{\scriptsize \IEEEauthorblockA{LIRMM, Univ. Montpellier, CNRS, Montpellier, France, Frederic.Comby@lirmm.fr}}\\
\and
\IEEEauthorblockN{Marc CHAUMONT}\\
{\scriptsize \IEEEauthorblockA{LIRMM, Univ Montpellier, CNRS, Univ. N\^imes, Montpellier, France, Marc.Chaumont@lirmm.fr}}}

\IEEEtitleabstractindextext{
\begin{abstract}
Image steganography aims to securely embed secret information into cover images. Until now, adaptive embedding algorithms such as S-UNIWARD or Mi-POD, were among the most secure and most often used methods for image steganography. With the arrival of deep learning and more specifically, Generative Adversarial Networks (GAN), new steganography techniques have appeared. Among them is the 3-player game approach, where three networks compete against each other.
In this paper, we propose three different architectures based on the 3-player game. The first architecture is proposed as a rigorous alternative to two recent publications. The second takes into account stego noise power. Finally, our third architecture enriches the second one with a better interaction between embedding and extracting networks. Our method achieves better results compared to existing works \cite{HayesNIPS2017_3players}, \cite{ZhuECCV2018_3players}, and paves the way for future research on this topic.
\end{abstract}

}

% make the title area
\maketitle

%\IEEEdisplaynontitleabstractindextext

%\ifCLASSOPTIONcompsoc
%\IEEEraisesectionheading{\section{Introduction}\label{sec:introduction}}
%\else
\section{Introduction}
\label{sec:introduction}
%\fi

In his paper, Simmons \cite{Simmons1983} formalized the reasoning framework for the steganography/steganalysis domain. It is defined as a {\it 3-player game}. The steganographs, usually named Alice and Bob, want to exchange a message without being suspected by a third-party. They need to create a secret communication channel in order to converse privately. So, they use a common medium, for example, an image, and dissimulate in this image a message. The steganalyst, usually named Eve, is observing the exchanges between Alice and Bob. If these exchanges are images, Eve has to check if they are natural (cover images) or if they hide a message (stego images). 

In the passive scenario, Eve does not modify the images \cite{Simmons1983}; Eve's role is only to make a binary decision, i.e. a two-class classification. Usually, in laboratory conditions \cite{Ker2013_RealWorld}, Eve has to be clairvoyant, meaning that she knows or has a good estimation of all the public parameters used by Alice and Bob, but she does not know their private parameters. These hypotheses about Eve's knowledge are close to the Kerckhoffs' principles \cite{Kerckhoffs1883} used in cryptography and are interesting when one wants to evaluate or compare the empirical security of steganographic embedding algorithms.

Modern embedding algorithms are adaptive, meaning that they take into account the content of the hosting medium (the cover) in order to better hide the message \cite{Holub2014_S-UNIWARD}, \cite{Holub2012_WOW}, \cite{Mi-POD}. 

Even if modern embedding approaches are the result of almost 20 years of research using codes and adaptivity, from a game theory point of view, these algorithms are qualified as {\it naive adaptive steganography} \cite{SchottleIH2012_Game}, \cite{SchottleTIFS2016_Game}. Indeed, when creating an embedding algorithm, the evolution of Eve's steganalysis strategy is not taken into account.

It is more interesting to propose an {\it optimal adaptive steganography} \cite{SchottleTIFS2016_Game}, also called {\it strategic adaptive steganography}. With such a steganography algorithm, pixels that would not have been modified by a naive approach have a chance to be modified. In other words, in a {\it strategic adaptive steganography}, the pixels' modification probability is set to ensure the Nash equilibrium in the cat-and-mouse game between Alice/Bob and Eve. 

{\it Strategic adaptive steganography} is a very nice concept, but trying to formalize it mathematically often requires simplifying assumptions which are far from modelling the practical reality. Another way to obtain a Nash equilibrium is to ``simulate'' the game. Alice can play the game alone (from her side and without interacting with Bob or Eve) by using {\it three algorithms}: the embedding algorithm, the extracting algorithm, and the steganalysis algorithm, which are competing against each other. We will name these algorithms {\it agents}; and more precisely, we will name {\it Agent-Alice} the embedding algorithm, {\it Agent-Bob} the extracting algorithm, and {\it Agent-Eve} the steganalysis algorithm, thus making a distinction with the {\it Human users} Alice (sender), Bob (receiver), and Eve (warden). Once, an equilibrium is achieved, Alice keeps her {\it strategic adaptive embedding} algorithm ({\it Agent-Alice}), and can send the extracting algorithm ({\it Agent-Bob}), or any equivalent information to Bob \footnote{This initial transfer from Alice to Bob is equivalent to the key exchange problem and will not be discussed in this paper.}. 

In reference to Simmons' formalization \cite{Simmons1983} and the computer science point of view (algorithms notion), we decided to name the approaches relying on the three agents, the {\it 3-player game} approaches.  The reader should nevertheless be aware that from a game theory point of view, there are only two teams that are competing (Alice plus Bob from one side, and Eve from the other side) in a zero-sum game. We believe that the ``{\it 3-player game}'' naming, better highlights the difference with the other families relying on adversarial approaches \cite{Chaumont2020}.

In the steganography domain, the pioneering approaches in order to find a {\it strategic equilibrium} date from 2011 and 2012, and were proposed in MOD \cite{MOD_Kodovsky} and in ASO \cite{ASO_Kouider} algorithms. Each of these two embedding-approaches iterates until a stopping criterion is reached between i) the embedding cost map update by Alice while requesting an Oracle (this is equivalent to an adversarial attack against a discriminant), and ii) the Oracle's update (update of the discriminant). 

In 2016, the authors of \cite{Abadi2016_AdversarialCrypto} proposed a cryptographic toy example: an encryption algorithm using three Neural Networks. The use of Neural Networks facilitates a {\it strategic equilibrium} since the problem is expressed as a min-max problem. Moreover, its optimization could be completed through the well-known back-propagation optimization process. Naturally, this {\it 3-player game} concept can be transposed in the steganography domain using deep learning. 

In December 2017 \cite{HayesNIPS2017_3players} and September 2018 \cite{ZhuECCV2018_3players}, two different teams from the machine learning community proposed, at NIPS 2017 and during ECCV 2018, to define {\it strategic embedding}, using 3 CNNs, iteratively updated, and playing the roles of the Agent-Alice, Agent-Bob, and Agent-Eve. These two papers provide an overview of the {\it 3-player game} concept, but the security notions and their evaluation are not treated correctly. When Eve is clairvoyant, both approaches are, in reality, very detectable.

More generally, the {\it 3-player game} approach belongs to one of the four GAN families, used in steganography \cite{Chaumont2020}.
These four families are the {\it no-modification/synthesis} SWE \cite{SWE_HU}, the {\it probability map generation} ASDL-GAN \cite{ASDL_JIWU}, the {\it adversarial} ADV-EMB \cite{ADV-EMB_TANG}, and finally, the {\it 3-player game}. In this paper, we only focus on the 3-player game approach. This approach requires the use of 3 CNNs and is totally different from the way the other families treat the problem. Therefore we will not compare our approach to the other families that are emerging. The philosophy of this paper is to clarify the {\it 3-player game} concept and propose practical solutions.    

In this paper, section \ref{sec:Stego3player} focuses on the steganography's main concept with the {\it 3-player game}. In section \ref{sec:related_work}, we recall the propositions given in \cite{HayesNIPS2017_3players} and \cite{ZhuECCV2018_3players}. 
In section \ref{sec:proposed_Architectures}, we present three architectures in order to resolve previous unsolved problems. In section \ref{sec:experiments}, we give some experimental results and their analysis. Finally, we conclude in section \ref{sec:Conclusion}.

\section{The 3-player game concept}
\label{sec:Stego3player}

\textbf{Notations:}
for this document, lowercase letters in bold are for vectors and matrices, lowercase letters in italic represent scalars.

``$\times$'' is used to separate the dimensions of multi-dimensional vectors and ``$\cdot$'' represents a multiplication.

Let \textbf{x} $\mathrm{\in }$ $\mathrm{\{0,...,255\}}^{w\times h}{}^{\ }$ be a cover matrix composed of \textit{w$\times$h} pixels, and \textbf{y} $\mathrm{\in }$ $\mathrm{\{0,...,255\}}^{w\times h}$ be a stego matrix with a size of \textit{w$\times$h} pixels generated by \textbf{Agent-Alice}. Let us further note \textbf{m} a secret binary message vector of \textit{m} bits that \textbf{Agent-Alice} wants to send to \textbf{Agent-Bob}, and \textbf{m'} the binary message extracted by \textbf{Agent-Bob}, where \textbf{m'} has the same length as \textbf{m}. Let\textbf{ k } be the shared key between \textbf{Agent-Alice} and \textbf{Agent-Bob,} where \textbf{ k } is a \textit{k}-sized binary vector. We note \textbf{z} $\mathrm{\in }$ $\mathrm{\{0,...,255\}}^{w\times h}$ an image with an unknown label. We use the notation \textit{l} for the image label where \textit{l}~$\mathrm{\in }$ $\mathrm{\{}$0,1$\mathrm{\}}$, \textit{l}=0 if \textbf{z} is a cover, and \textit{l} =1 if \textbf{z} is a stego. 

%%%%%%%%%%%%%%%%%%%%%%%%%%%
\subsection{General concept}
\begin{figure*}[htb]
\includegraphics[width=7.6in,height=1.8in, keepaspectratio=false]{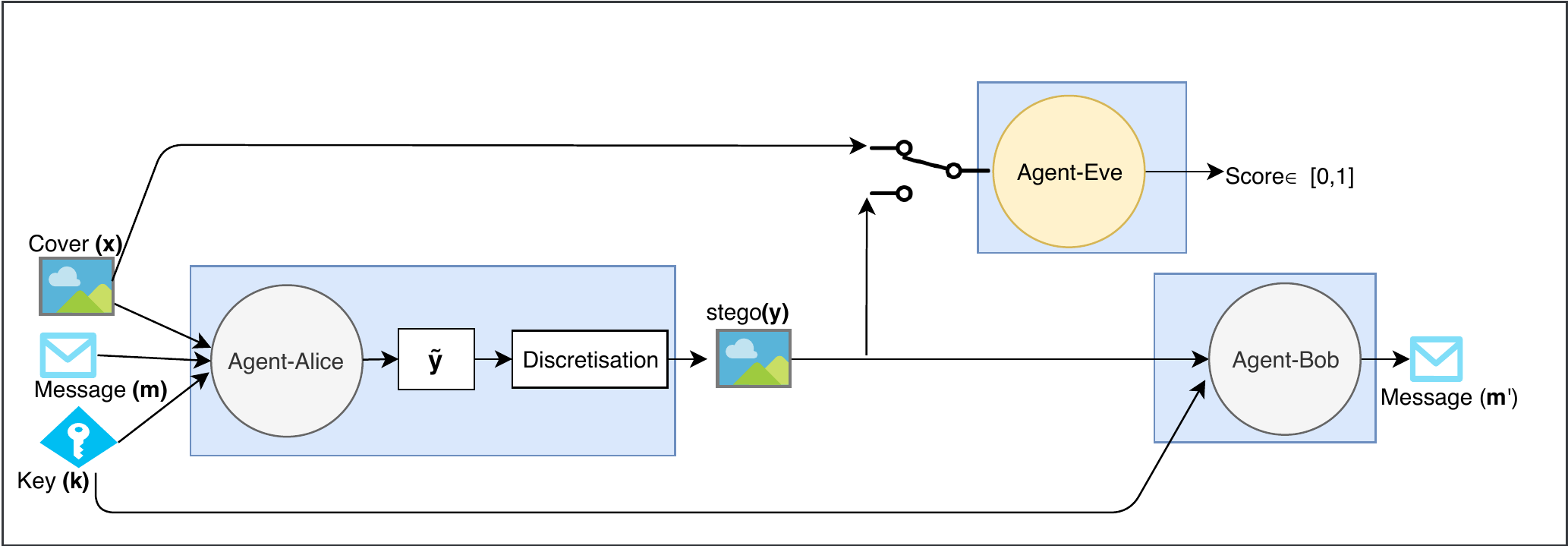}
\caption{The overall architecture of the {\it 3-player game}.}
\label{fig:The overall architecture}
\end{figure*}

This part of the paper introduces the general concept of the {\it 3-player game} and describes the role of each agent. The {\it 3-player game}-based steganographic system illustrated in Fig.~\ref{fig:The overall architecture} is composed of three neural networks. These networks represent the three agents: Agent-Alice, Agent-Bob, and Agent-Eve.

The system's input consists of a cover image \textbf{x}, a secret message\textbf{ m } and a key\textbf{ k}. These inputs are first introduced to Agent-Alice's network that generates a non-discrete-stego $\mathbf{\tilde{y}}\mathrm{\in }$ $\mathrm{\mathbb{R}}^{w\times h}$. Then the discretization module receives $\mathbf{\tilde{y}}$ and generates \textbf{y} a stego image with discrete values. \textbf{y} is then given to both Agent-Bob and Agent-Eve.

Agent-Bob tries to recover the secret message \textbf{m} from the stego \textbf{y} using the shared secret key \textbf{k}. Agent-Bob inputs (\textbf{y}, and \textbf{k}) and goes through a set of layers and  mathematical operations; the extracted message \textbf{m'} is then generated.

Agent-Eve\ receives an image \textbf{z}, and returns a probability score of \textbf{z}'s membership to the cover or stego classes. 

%%%%%%%%%%%%%%%%%%%%%%%%%%%%%%%%%%%%%%%%%%%
\subsection{Three agent's losses}

The objective of the steganographic system shown in Fig.~\ref{fig:The overall architecture} and described previously is to learn a model, so Agent-Alice can generate a stego  \textbf{y} by embedding the secret message \textbf{m} within the cover \textbf{x}, and then secretly communicate it to Agent-Bob. Given this objective a loss function is given to each agent:

\textbf{Agent-Eve's loss: }
\noindent Let an image \textbf{z} =(z$_{ij})^{w\times h}$ whose label \textit{l} is unknown to Agent-Eve. Agent-Eve is modelled by a function $Agent\text{-}Eve$: \textbf{z} $\rightarrow$ [0, 1], which takes the image \textbf{z} and returns a real score between 0 and 1, such that 0 corresponds to a cover and 1 corresponds to a stego.

Agent-Eve's general loss consists in minimizing the distance between the label \textit{l} and Agent-Eve's prediction:
\begin{gather}
\mathcal{L}_{Eve} = dist(\textit{l}- Agent\text{-}Eve (\textbf{z})).
\label{eq:Eve}
\end{gather}

The distance used for Agent-Eve's loss is usually the cross-entropy distance; thus the loss of Eq.~\ref{eq:Eve} is given as:

\begin{align}
\mathcal{L}{}_{Eve} = &- \textit{l}\cdot log (Agent\text{-}Eve (\textbf{z})) \nonumber \\
  & - (1\textit{ - l})\cdot log (1 - Agent\text{-}Eve (\textbf{z})).
\label{eq:Eve-cross}
\end{align}

\textbf{Agent-Bob's loss: }
\noindent Agent-Bob attempts to reconstruct the secret message \textbf{m} from the received image \textbf{y} using the key \textbf{k}. The reconstructed message \textbf{m'} should be equal to \textbf{m} (\textbf{m} =\textbf{ m'}). To this end, Agent-Bob's loss consists in minimizing a distance between \textbf{m'} and \textbf{m} (usually a L2 distance):
\begin{gather}
\mathcal{L}_{Bob} = dist(\textbf{m},\textbf{m'}).
\label{eq:Bob}
\end{gather}

\textbf{Agent-Alice's loss: }
\noindent Agent-Alice's objectives are multiple. The first is to generate a stego image \textbf{y} that is close enough to the cover \textbf{x}. The second is to allow Agent-Bob to reconstruct the secret message \textbf{m} correctly from the stego image. The third is that Agent-Eve's accuracy should not be better than a random guess whether a given image \textbf{z} is a cover or stego (50-50 chance of making the correct guess). The loss of Agent-Alice is then the weighted sum of three terms: $\mathcal{L}_{bob}$, $\mathcal{L}_{Eve}$, and \textit{dist}(\textbf{x,y}) the distance calculated between \textbf{x} and \textbf{y}, where all coefficients $\lambda_A$, $\lambda_B$, $\lambda_E$ belongs to [0,1] and sum to one in order to adjust the contribution of each term to the loss of Agent-Alice:
\begin{gather}
    \mathcal{L}{}_{Alice} = \lambda_A \cdot dist(\textbf{x,y})+ \lambda_B  \cdot\mathcal{L}{}_{Bob}-\lambda_E \cdot\mathcal{L}{}_{Eve}.
    \label{eq:Alice}
\end{gather} 

Note that pixel values from {\bf x} and {\bf y} are all normalized by a division by 255. So, each of the three terms has similar value ranges, which is a practical requirement in an optimization process (see Fig.~\ref{fig:loss-curves}).

\subsection{Training process}
Now that we have presented the general concept of the {\it 3-player game} and the loss for each agent, we present the algorithm used for the training process.

\begin{algorithm}
\SetAlgoLined
\KwResult{{\it stegos, extracted\_messages}}
\KwData{{\it covers-list, messages-list, keys-list}}
 \While{not converge OR $loop  \leq \text{\it max-iter}$}{
     \tcp{Alice and Bob learning}
   {\For{$iter\_team1 \leq it1$}{
     get\_batch ({\it covers\_list, messages\_list,keys\_list,batch\_size})\;
     forward-propagation ({\it covers, messages,keys})\;
     update\_Agent-Bob ($\mathcal{L}_{Bob}$);\\
     update\_Agent-Alice ($\mathcal{L}_{Alice}$);\\
   }}
     \tcp{Eve learning}
   {\For{$iter\_team2 \leq it2$}{
     get\_batch ({\it covers\_list, stegos\_list})\;
     forward-propagation ({\it covers, stegos})\;
     update\_Agent-Eve ($\mathcal{L}_{Eve}$);
   }}
 }
 \caption{\textit{3-player game} training process}
\label{Algo:3players}
\end{algorithm}

As shown in \textbf{Algorithm~\ref{Algo:3players}} on line 1, the global system is trained at the maximum for \textit{max-iter} ``{\it loop}''. In each loop, the learning is completed sequentially by first, the team Agent-Bob and Agent-Alice (line 2) and then, the Agent-Eve (line 8). Note that there is a high number of loops in order to reach an equilibrium. Also, note that inside each loop there is also a certain number of back-propagation iterations for each agent. 

Therefore, for the learning of Agent-Bob and Agent-Alice (lines 2 to 7), there are {\it it1} iterations (line 2). For an iteration, we load a mini-batch of {\it cover images}, {\it secret messages}, and {\it keys} (line 3), we forward-propagate all the cover images on Agent-Bob's and Agent-Alice's networks (line 4), and then we update Agent-Bob and Agent-Alice by minimizing the $\mathcal{L}_{Bob}$ and $\mathcal{L}_{Alice}$ thanks to the stochastic gradient descent (lines 5 and 6). During this learning phase, the weights of Agent-Eve are fixed.

The learning of Agent-Eve (lines 8 to 12) is similar to the learning of Agent-Bob and Agent-Alice. There are {\it it2} iterations (line 8). For an iteration, we load a mini-batch of {\it cover images} and {\it stego images} (line 9), we forward-propagate all the cover and stego images on Agent-Eve's network  (line 10), and then we update Agent-Eve by minimizing the $\mathcal{L}_{Eve}$ thanks to the stochastic gradient descent (lines 11). 

When the equilibrium is reached, the last agent playing the game will not change its strategy. So, the fact to be the last player, i.e. the last learning agent, will not impact the performances of the other agents.

%%%%%%%%%%%%%%%%%%%%%%%%%%%%%%%%%%%%
\section{Related work}
\label{sec:related_work}
In this section, we recall the architectures of GSIVAT~\cite{HayesNIPS2017_3players} and HiDDen~\cite{ZhuECCV2018_3players}. These two architectures were basically proposed for steganography purposes. They belong to the \textit{3-player game} family.
\subsection{Generating Steganographic Images Via Adversarial Training (GSIVAT)} 
\begin{figure*}% height=3.3in,
\begin{center}\includegraphics*[width=7.2in, height=3.in, keepaspectratio=false]{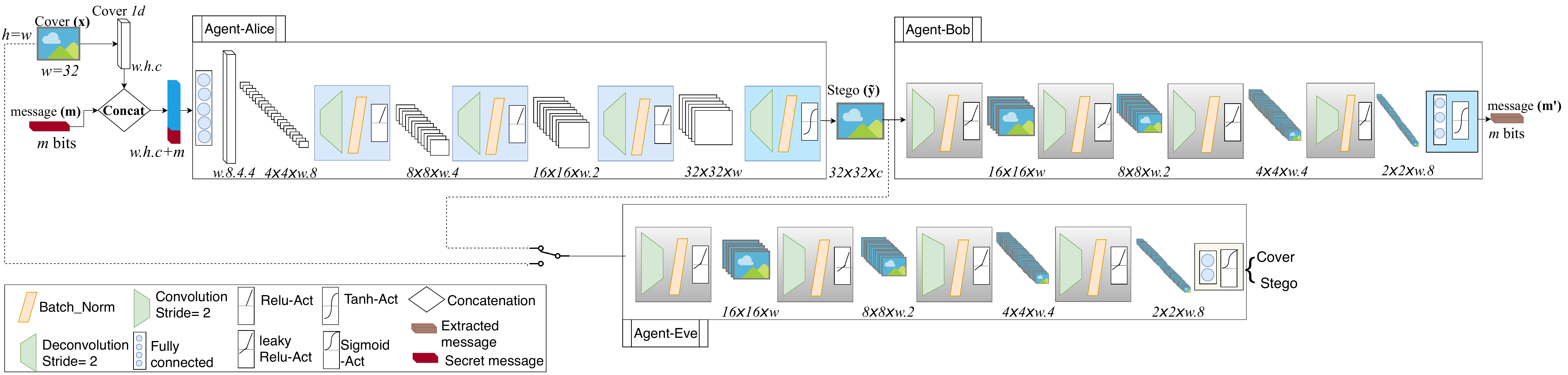}\end{center}
\caption{GSIVAT \cite{HayesNIPS2017_3players} architecture.}
\label{fig:Haydes architecture}
\end{figure*}

In \cite{HayesNIPS2017_3players}, the authors propose a steganographic system (GSIVAT) composed of three neural networks, each one representing one agent (Agent-Alice, Agent-Bob, Agent-Eve). We provide an overview of the GSIVAT architecture in Fig.~\ref{fig:Haydes architecture}.

Their system's input are a cover image \textbf{x}, a \textit{3D} vector, whose size is \textit{w$\times$h$\times$c} (where \textit{c} is the channel number) and a secret message \textbf{m} of \textit{m} bits. \textbf{x} is flattened to a 1D vector and concatenated with \textbf{m}, the resulting vector size is \textit{w$\cdot$h$\cdot$c+m}. This vector is then fed to Agent-Alice.

\textbf{Agent-Alice} is composed of one fully connected layer (FC) and four deConv-BN-Act blocks (see Fig.~\ref{fig:Haydes architecture}). Each block is the combination of a deconvolution (deConv), a batch normalization (BN) and one activation function (Act) layer (ReLU, leaky-ReLU, sigmoid, tanh).

The number of neurons in the fully connected (FC) layer is the multiplication result of \textit{w}$\cdot$8$\cdot$4$\cdot$4. The output of the FC is reshaped to \textit{w}$\cdot$8 of 4$\times$4 sized feature-maps. Then, these feature-maps feed four \textit{deConv-BN-Act blocks}. After each block, the size of the feature-maps is up-sampled by a factor of two, and the channel number is reduced by a factor of two. In the end, an image of the same size as \textbf{x} is obtained.

Agent-Alice outputs the stego image $\mathbf{\tilde{y}}\mathrm{\in }$ $\mathrm{\mathbb{R}}^{w\times h}$ which is fed to both Agent-Bob and Agent-Eve. The distance between \textbf{x} and $\mathbf{\tilde{y}}$ (dist(\textbf{x,y}) in Eq.~\ref{eq:Alice}) is the L2 distance.

\textbf{Agent-Bob}'s architecture is depicted in Fig.~\ref{fig:Haydes architecture}. It consists of four \textit{Conv-BN-Act} blocks and one FC layer. Each block is made up of a fractionally-strided (stride=2) convolution layer (Conv), a batch normalization layer and an activation function ``Leaky ReLU'' \cite{LReLU}. The FC layer is made up of \textit{m} neurons followed by a tanh activation function.  

Agent-Bob outputs the extracted message \textbf{m'}. The L2 distance is used to compute the loss previously given in Eq.~\ref{eq:Bob}.

\textbf{Agent-Eve} takes as input image \textbf{z}, and outputs a score determining the label of the input (cover/stego). The architectures of Agent-Eve and Agent-Bob are similar, except for the fully connected layer having only two neurons, and the activation function replaced by a sigmoid. The sigmoid cross-entropy loss is used for Eve's loss (see Eq.~\ref{eq:Eve-cross}).
%%%%%%%%%%%%%%%%%%%%%%%%%%%%%%%%%%%%%%%%%%
\begin{figure*}[htb]%height=3in,
\includegraphics[width=7in, height=2.3in, keepaspectratio=False]{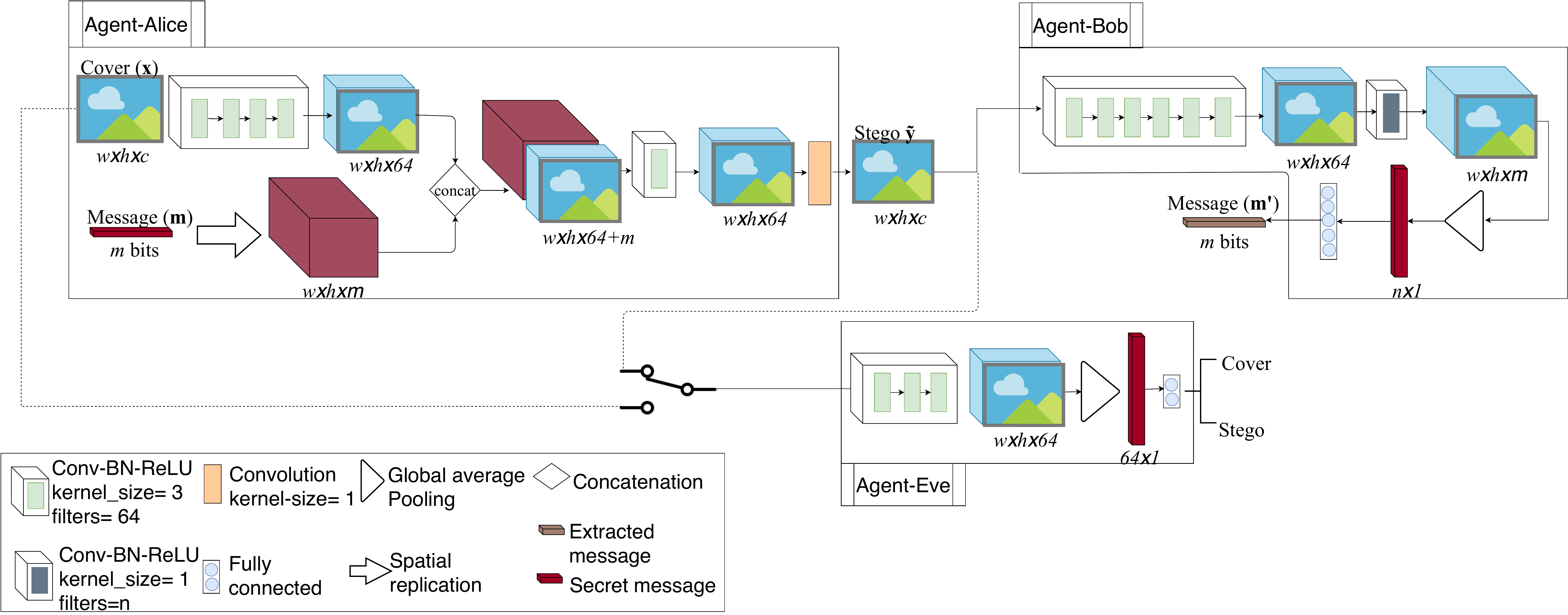}
\caption{HiDDeN architecture.}
\label{fig:HiDDeN architecture}
\end{figure*}
\subsection{HiDDeN: Hiding Data With Deep Networks}

In \cite{ZhuECCV2018_3players}, the authors propose a model for image steganography, but also for watermarking (called HiDDen). We will only discuss the architecture of their steganographic system.

Their system takes as input a \textit{w$\times$h$\times$c}-sized cover image \textbf{x} and a \textit{m}-length secret message \textbf{m}. The system is composed of three neural networks as illustrated in Fig.~\ref{fig:HiDDeN architecture}. The network's structure is generally composed of blocks called \textit{Conv-BN-ReLU} which stands for the combination of Convolution, Batch Normalization and a ReLU activation function.

\textbf{Agent-Alice} is composed of five \textit{Conv-BN-ReLU} blocks and one convolution layer with a kernel size of 1$\times$1. Firstly, the cover image \textbf{x} goes through four \textit{Conv-BN-ReLU} blocks to obtain an intermediate representation image $\mathbf{\tilde{x}}$. Firstly, the message \textbf{m} is replicated so that the resulting size is \textit{w$\times$h$\times$m}. Secondly, $\mathbf{\tilde{x}}$ is concatenated with the replicated message and fed to another \textit{Conv-BN-ReLU} block with 64 output filters. A final convolution layer with a 1$\mathrm{\times}$1 kernel is used to generate the stego image $\mathbf{\tilde{y}}\mathrm{\in }$ $\mathrm{\mathbb{R}}^{w\times h}$. The loss of Agent-Alice is calculated using the loss of Eq.~\ref{eq:Alice}.

\textbf{Agent-Bob} is composed of seven \textit{Conv-BN-ReLU} blocks, followed by global spatial average pooling (to produce a vector with the same size as the message). Then, a single fully connected layer ends the architecture, as shown in Fig.~\ref{fig:HiDDeN architecture}. Agent-Bob receives the stego image $\mathbf{\tilde{y}}$, and produces the predicted message \textbf{m'}. The distance used in Agent-Bob's loss of Eq.~\ref{eq:Bob} is the l2 distance. 

\textbf{Agent-Eve} takes an image as input, and outputs a score indicating whether the given image is a cover or stego. Agent-Eve has an architecture similar to Agent-Bob, but only has three \textit{Conv-BN-ReLU} blocks instead of seven. The last layer is a FC layer with an output size of two units (see Fig.~\ref{fig:HiDDeN architecture}). The authors adopt the use of the cross-entropy loss presented in Eq.~\ref{eq:Eve-cross} for Agent-Eve.
%%%%%%%%%%%%%%%%%%%%%%%%%%%%%%%%%%%%%%%%%%%%%%%%%%%%%%%%%%%%%%%%%%%%%%%%%%%%%%%%%%
\subsection{Discussion}
\label{sec:discussion}
The two papers presented previously \cite{HayesNIPS2017_3players} \cite{ZhuECCV2018_3players} offer some interesting ideas, but there are flaws in both. 

Firstly, neither of the two approaches use a shared secret key during the embedding/extracting process. The authors of  \cite{HayesNIPS2017_3players} and \cite{ZhuECCV2018_3players} suppose that the information about model weights, architecture, and the set of images used for training is shared between Agent-Alice and Agent-Bob. According to the authors, this shared information can be considered as the secret key. Besides the fact that such a hypothesis is heavy in size (almost 70 Mb sent to Agent-Bob \cite{HayesNIPS2017_3players}), it is also the equivalent of performing an embedding process with the same \textit{secret-key}. Such an embedding process is highly discouraged in steganography as it leads to very easy detectability \cite{Pibre2016}.

Secondly, there is no discretization module for the generated images (Agent-Alice provides $\mathbf{\tilde{y}}$ instead of \textbf{y}). In a real-world situation \cite{Ker2013_RealWorld}, Bob receives an image whose values are defined in $\mathrm{\{}$1,..., 255$\mathrm{\}}$, and has to extract the secret message. In \cite{HayesNIPS2017_3players} and \cite{ZhuECCV2018_3players}, Agent-Alice generates real-valued images i.e. {\bf not} discrete-value images, and these images are fed to Agent-Bob. This makes both Agent-Alice and Agent-Bob useless in practice.  Alice needs to provide Bob with images that are not suspicious, meaning images with discrete values. Indeed, images have to be formatted in PGM, or any lossless compressed image format. Note that if Alice decides to round the real-value images (generated by Agent-Alice) in order to discretize them in $\mathrm{\{}$1,..., 255$\mathrm{\}}$, Bob, when using Agent-Bob algorithm, will not extract the message correctly, since Agent-Bob has been built for real-value images\footnote{We observed this phenomenon during our experiments.}.

Thirdly, the computation load is a serious issue that we need to take into account when working with deep learning. GSIVAT authors \cite{HayesNIPS2017_3players} worked on 32$\times$32-sized images while Hidden authors \cite{ZhuECCV2018_3players} used 16$\times$16-sized patches. This limitation for the size of the images is due to the use of the FC layers, which introduce expensive memory and computation costs.
The authors suggest that working on large images could be completed by treating bigger images with a separate treatment for each part of the image.
This is a bad idea since statistical traces may be found at the block boundaries and would lead to an easily detectable embedding scheme (See for example, the discussion in section 4.2 of \cite{Fridrich2009_Book}, or the dependencies preservation between blocks in JPEG steganography \cite{Taburet2019_Natural_JPEG}).

Finally, note that in both of these papers, the experimental steganalysis is performed with the algorithm ATS \cite{ATS2015} proposed in 2015. This algorithm is basically designed to handle the cover-source mismatch problem, which is definitely not the appropriate scenario to evaluate the empirical security of an embedding algorithm (especially when it is a \textit{strategic embedding} algorithm). Indeed, ATS is based on the assumption of \textit{constant noise direction} in the embedding space, which may not be true for \textit{a strategic adaptive} algorithms. The empirical security is probably undervalued when compared to an  Ensemble Classifier/Rich Model (EC+RM) \cite{Kodovsky2012-EnsembleClassifiers, Fridrich2012_Rich}, Yedroudj- Net \cite{YedroudjCNN}, ReST-Net \cite{ReST-Net}, or SRNet \cite{SR-Net}. In addition, these four steganalysis algorithms represent the current state-of-the-art in steganalysis, so their use makes more sense.

\section{Our steganographic system's Architecture}
\label{sec:proposed_Architectures}

In this paper, we propose a new \textit{strategic adaptive steganography} system based on the \textit{3-player game} concept. We are using an embedding algorithm (Agent-Alice) and an extracting algorithm (Agent-Bob) which functions in practice. 
We therefore:
 
\begin{enumerate}
\item Integrate a stego-key for the input of Agent-Alice and Agent-Bob. With two different stego-keys, Agent-Alice will generate two different stego images. Alice knows that she must change the stego-key very often if she doesn't want to be caught \cite{Pibre2016}. By extension, knowing that it is easier to break a system that always uses the same key, it is important to integrate a stego key in the input of Agent-Alice and Agent-Bob, in order to avoid the counter-productivity that a unique key could have on the convergence of Agent-Alice and Agent-Bob facing Agent-Eve. This argument is not considered at all in \cite{HayesNIPS2017_3players} and \cite{ZhuECCV2018_3players} and can be a major flaw in their performances.
\item Handle the problem of discretization in order for Alice to be able to send to Bob, through  e-mail, memory stick, cloud storage, an image in a non-suspicious standard format. (\cite{HayesNIPS2017_3players} and \cite{ZhuECCV2018_3players} do not deal with this fundamental issue).
\item Guarantee a scalable (in memory and in computation) solution thanks to an architecture that consists of only convolutions. This way, it can deal with image dimensions usually used in deep-learning and steganalysis by deep-learning in academic experiments ($255\times 255$ or $512\times512$) \cite{Chaumont2020}. The convolutional architectures also allow deeper networks to deal with harder problems modelization. GSIVAT \cite{HayesNIPS2017_3players} works with $32\times32$ images and HiDDeN \cite{ZhuECCV2018_3players} works with $16\times16$ images and they both use very small networks.
\end{enumerate}

Additionally, our approach offers two interesting properties. Firstly, it is ``bit-rate adaptive''. Indeed, there is no need to re-train the system each time we change the bit rate, i.e. each time the message size is different (this is not the case for \cite{HayesNIPS2017_3players} and \cite{ZhuECCV2018_3players}). Secondly, we adopt a strong steganalyst for Agent-Eve, which benefits from better security, but it is not the most up-to-date steganalysis.

We propose three different architectures for our steganographic system. These architectures illustrate three different solutions going from a basic one, to a more appropriate solution. On these three architectures, Agent-Eve's remains the same, while the design of Agent-Alice and Agent-Bob's changes.
The first architecture is presented to illustrate the use of a secret shared key during embedding. The second architecture has been conceived to reduce the power of noise introduced by embedding the hidden message into the cover image. Finally, the third architecture tries to improve the performances of message extraction while linking Agent-Alice and Agent-Bob's behaviour.

\subsection{The training process}
Looking at the three proposed architecture, the training procedure of the system remains the same. We alternate the training between the three agents, Agent-Alice, Agent-Bob and Agent-Eve, where Agent-Alice and Agent-Bob are trained jointly as a single network, and Agent-Eve is trained separately.

First, Agent-Bob and Agent-Alice are trained on a fixed number of mini-batches using the two models of Agent-Alice and Agent-Bob saved previously. See Algorithm.~\ref{Algo:3players}. This training process is repeated for several loops, until all losses tend to be constant. 

\subsection{The proposed architecture of the Agent-Eve}
Agent-Eve tries to guide both Agent-Alice and Agent-Bob through the process of learning a \textit{strategic adaptive embedding} algorithm. If Agent-Eve is weak, the 3-agent system falls down. Indeed, Agent-Alice and Agent-Bob will no longer search for better solutions as Agent-Eve cannot cope with their evolution. To this end, it is essential to adopt a strong steganalyzer.

In 2018, the best spatial steganalyst was, Yedroudj-Net\cite{YedroudjCNN}\cite{Yedroudj_EI18} (first published in January 2018),  ResT-Net\cite{ReST-Net} (published in March 2018) and more recently SRNet \cite{SR-Net} (published in September 2018). Among these networks, Yedroudj-Net is the shallowest network, with six convolution layers compared to 25 layers for SRNet, and 30 layers for ReST-Net (3 sub-networks each containing 10 layers). Besides its affordable size, training Yedroudj-Net does not require the use of any tricks that could increase the computational time. This network is therefore well adapted to Agent-Eve, especially knowing that \textit{the 3-player approach} takes a lot of time before it converges to a good solution. Additionally, Yedroudj-Net can easily be improved in the future, if required \cite{Zhu-Net}.

\begin{figure*}[!htb]
\includegraphics[width=7in, height=2.in, keepaspectratio=false]{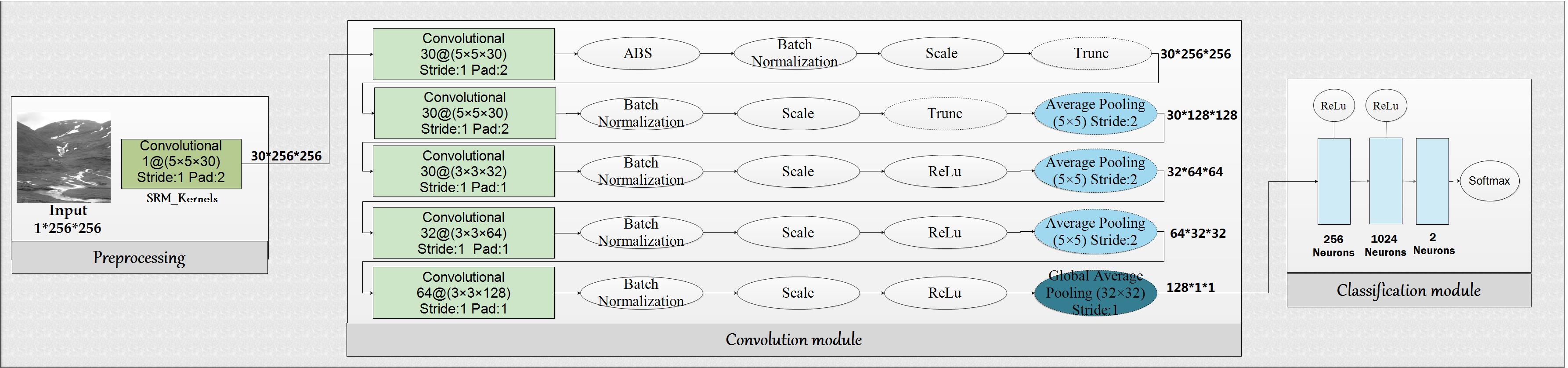}
\caption{The overall architecture of Agent-Eve \cite{YedroudjCNN}.}
\label{fig:Eve's architecture}
\end{figure*}

Yedroudj-Net architecture \cite{YedroudjCNN} is presented in Fig.~\ref{fig:Eve's architecture}. It is composed of 7 blocks: a pre-processing block, five convolutional blocks, and a fully connected block made of three fully connected layers followed by a softmax \footnote {For more details on Yedroudj-Net, the reader can view the online code at \href{http://www.lirmm.fr/~chaumont/Yedroudj-Net.html}{www.lirmm.fr/$\mathrm{\sim}$chaumont/Yedroudj-Net.html} }.

Agent-Eve's network (Yeroudj-Net) is trained by minimizing the loss given in Eq.~\ref{eq:Eve-cross}.
%%%%%%%%%%%%%%%

\subsection{First-Architecture}

The first architecture is similar in spirit to the previous approaches of existing literature except that it contains only convolutional layers, and integrates a stego-key.

Agent-Alice's network receives a \textit{m}-length secret message \textbf{m}, a key \textbf{k} of \textit{k} bits, and a cover \textbf{x}. in order to concatenate the cover \textbf{x} with the message \textbf{m}, both should have the same size. So, we use the key \textbf{k} to spread out the secret message \textbf{m} in a matrix noted as $\mathbf{s^{(m)}}\mathrm{\in }\mathrm{\{0,1\}}{}^{w\times h}$ that is filled with zeros and has the same size as our cover image.\\
The spreading of \textbf{m} in $\mathbf{s^{(m)}}$ is obtained by using a pseudo random number generator (PRNG) seeded by the key \textbf{k}. The PRNG sequentially picks a bit of \textbf{m} and fills a \textit{non-used} position in the $\mathbf{s^{(m)}}$ matrix. The filled positions define the binary mask ${\bf \Omega} \mathrm{\in }\mathrm{\{0,1\}}{}^{w\times h}$. ${\bf \Omega}$ therefore contains exactly \textit{m} ones.

Note that with the knowledge of \textbf{k}, and the index of a bit $m_i$ in our message \textbf{m} with $i\mathrm{\in }\mathrm{\{1,...,m\}}$, we can deduce the position $(u,v) \mathrm{\in }\mathrm{\{0,..\textit{w}\}\times \{0,..\textit{h}\}}$ where this bit is stored in $\mathbf{s^{(m)}}$, and inversely from a position  $(u,v) \mathrm{\in }\mathrm{\{0,..\textit{w}\}\times \{0,..\textit{h}\}}$, we can deduce the bit  $m_i$ with $i\mathrm{\in }\mathrm{\{1,...,m\}}$ of the message \textbf{m}.

\begin{figure*}[htb]% height=3in,
\includegraphics[width=7in,height=3in, keepaspectratio=false]{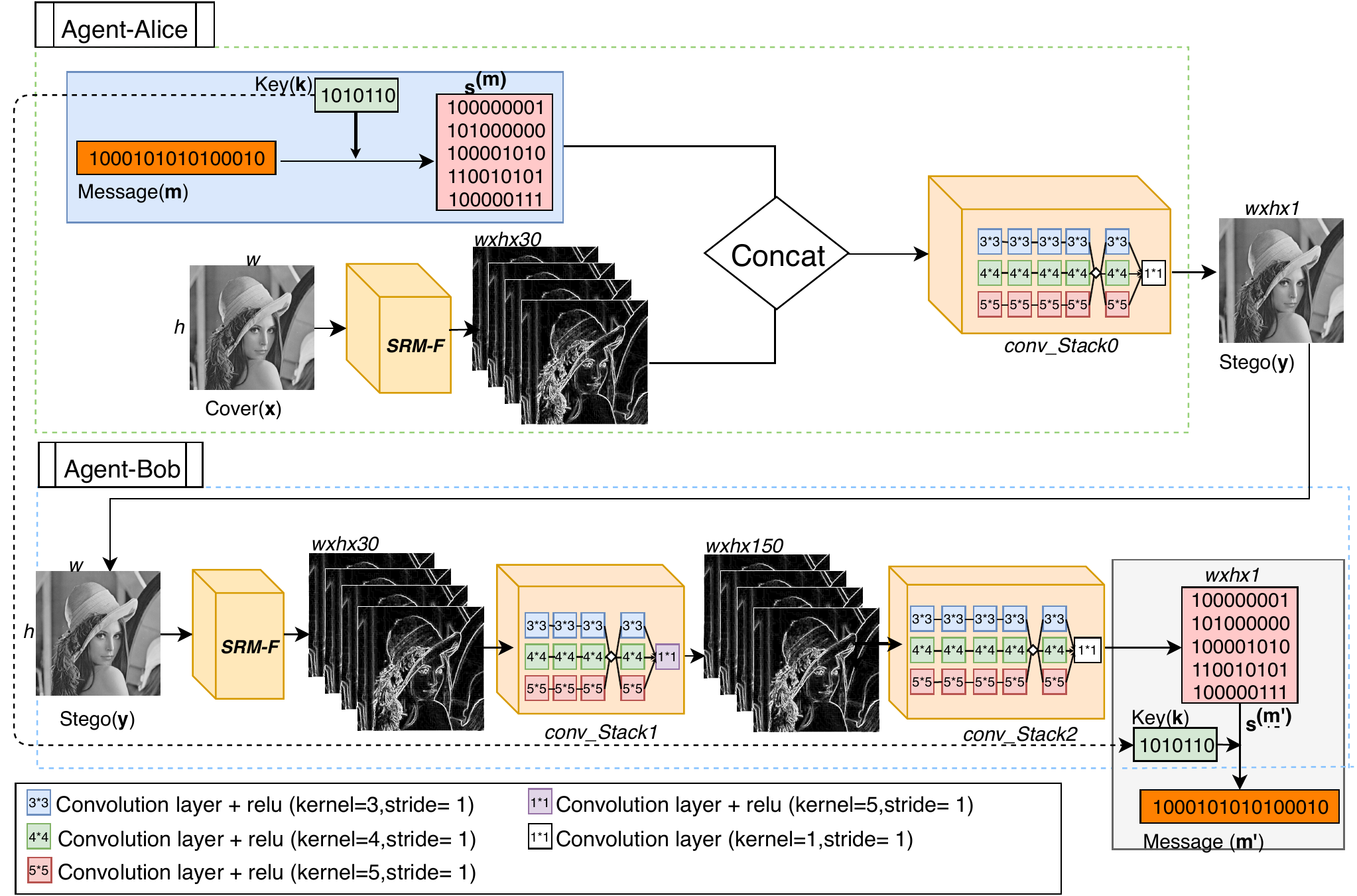}
\caption{Agent-Alice and Agent-Bob with the first architecture.}
\label{fig:first architecture}
\end{figure*}

The cover image \textbf{x} is fed to a convolution layer called ${\textit{SRM-F}}$. Its weights are initialized with the 30-basic high-pass filters of SRM \cite{Fridrich2012_Rich}, similarly to Yedroudj-Net \cite{YedroudjCNN}. The output is then concatenated with $\mathbf{s^{(m)}}$ and fed to \textit{conv\_Stack0} which is composed of a set of convolution layers. The output is a $w\times h$ image which represents the stego image \textbf{y} (see Fig.~\ref{fig:first architecture}).

\noindent 
Agent-Bob uses the stego \textbf{y} and the key \textbf{k} to output the predicted message \textbf{m'}, a vector of \textit{m} bits. First, the stego image passes through a convolution layer similar to \textit{SRM-F}. Next, the obtained feature-maps from \textit{SRM-F} go through another set of convolutions in two \textit{conv\_Stack1, conv\_Stack2} (Note that \textit{conv\_Stack2} shares the same construction as \textit{conv\_Stack0}, but not the same number of features-maps for each layer) . The extracted message  \textbf{m'} is then obtained (see Fig.~\ref{fig:first architecture}).

The back-propagation is done by minimizing both losses of Agent-Alice and Agent-Bob. For Agent-Bob's loss (see Eq.~\ref{eq:Bob}), the Mean Square Error (MSE) distance between \textbf{m} and \textbf{m'} is used. It is written as: 
\begin{gather}
   \mathcal{L}_{Bob}=\left( \mathrm{\sum}^{m}_{i=1}(m_i\mathrm{-} m'_i)^{2}\right)/m\, , 
\label{eq:Bob_L2}
\end{gather}which is equivalent to:
\begin{gather}
   \mathcal{L}_{Bob}=( {(\mathbf{s^{(m)}}-\mathbf{s^{(m')}}) \odot \Omega)^{2} /m \, ,} 
     \label{eq:Bob_mask}
\end{gather}with $\odot$ the point-wise operation.

Agent-Alice uses the loss presented in Eq.~\ref{eq:Alice} with \textit{dist}(\textbf{x,y}) calculated as:
\begin{gather}
    dist(\textbf{x},\textbf{y})=\left ( {\sum}^{w}_{i=1}{\sum}^{h}_{j=1}(x_{ij}\mathrm{-} y_{ij})^{2}\right)/w.h \, ,
\label{eq:dist}
\end{gather}
and so Agent-Alice's loss is:
\begin{gather}
\mathcal{L}_{Alice}=\lambda_A \cdot dist(\textbf{x},\textbf{y}) + \lambda_B\cdot \mathcal{L}_{Bob}-\lambda_E\cdot \mathcal{L}_{Eve}
\label{eq:Alice-loss1}
\end{gather}
where $\lambda_A$, $\lambda_B$, $\lambda_E$ $\mathrm{\in }$ belongs to [0,1] and sum to one in order to adjust the contribution of each term to the loss.
In this architecture, the magnitude of modifications is not restricted and therefore the stego noise power is too strong. To impose stego noise power restriction, we propose a second architecture. 

\subsection{Second-Architecture (noise power reduction)}
\label{Second-Architecure}
\begin{figure*}[htb]
\includegraphics[width=7in,height=3in, keepaspectratio=false]{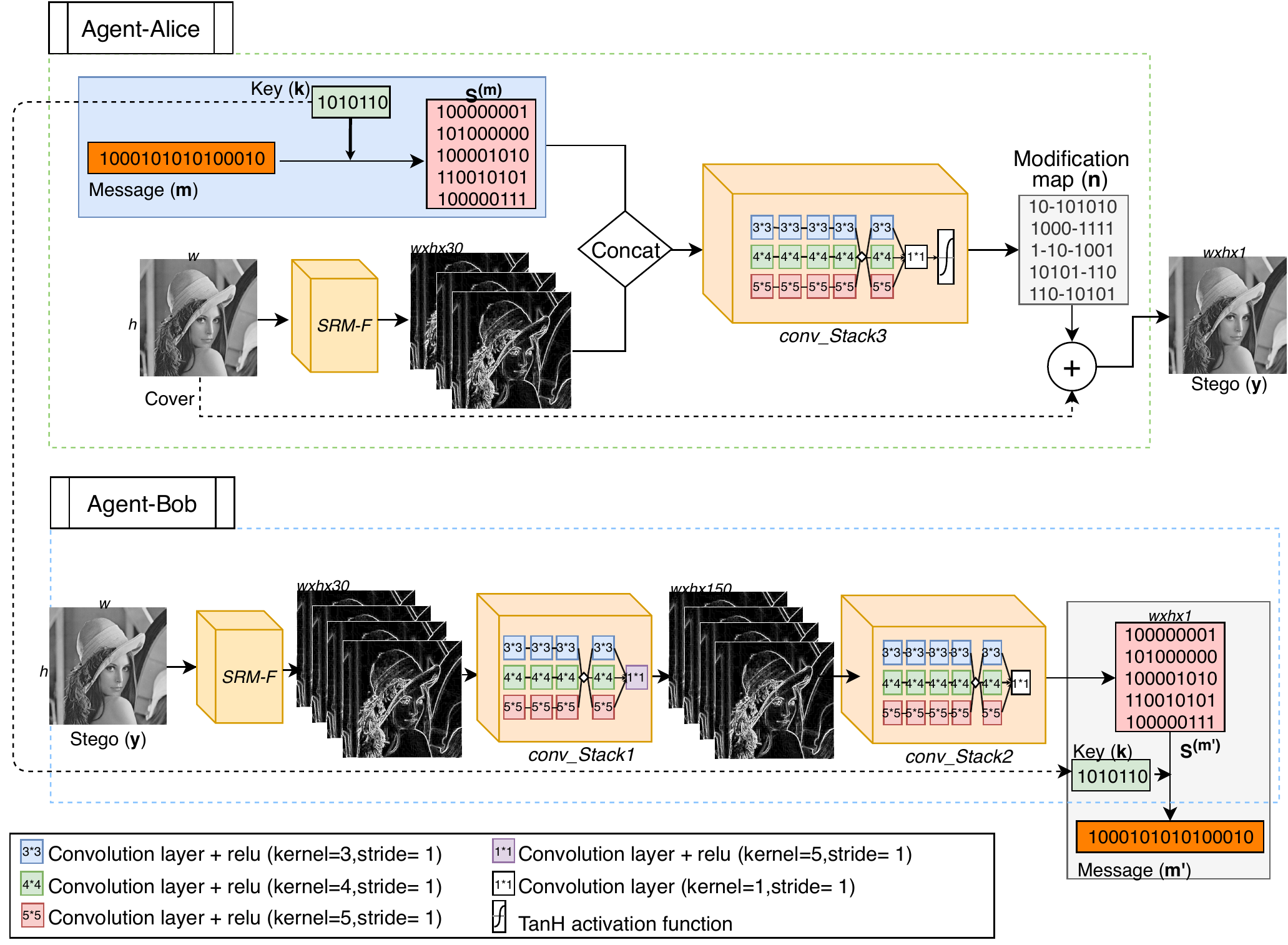}
\caption{Agent-Alice and Agent-Bob with the second architecture.}
\label{fig:second architecture}
\end{figure*}

The second architecture improves the first one by imposing a stronger restriction on the magnitude of modification of stego noise. We force Agent-Alice to make the least amount of changes to the cover whilst still allowing Agent-Bob to retrieve the secret message correctly. Within this architecture, see Fig.~\ref{fig:second architecture}, Agent-Bob's design remains the same as in the previous architecture. However, the architecture of Agent-Alice has changed. Instead of letting the network decide the intensity of modification for each pixel in the cover image, it is restricted to a ternary modification; the stego noise values are $\mathrm{\{}$-1, 0, 1$\mathrm{\}}$. More precisely, during the first iterations, values are in the range of $[-1, 1]$ and belong to $\mathbb{R}$. But at the end of iterations, a discretization is completed in order to have only three discrete values $\mathrm{\{}$-1, 0, 1$\mathrm{\}}$.

Said differently, Agent-Alice generates a modification map $\textbf{n}\mathrm{\in}\mathrm{\{}$-1, 0, 1$\mathrm{\}}$ which is then added to the cover image \textbf{x} to generate the stego \textbf{y} directly (see Fig.~\ref{fig:second architecture}). The generation of \textbf{n} is performed thanks to the resulting feature maps of $\textit{SRM-F}$, and the spread message $\mathbf{s^{(m)}}$. These are both concatenated and fed to \textit{conv\_Stack3}; the output of the latter is \textbf{n} in the range of $[-1, 1]$ thanks to a TanH activation function, and is one of the three discrete values $\mathrm{\{}$-1, 0, 1$\mathrm{\}}$ once the discretization is activated.

Agent-Bob's network loss remains the same as in Eq.~\ref{eq:Bob_L2}. For Agent-Alice's loss, we calculate the mean of the absolute values of the modification maps \textbf{n} (which is equivalent to the MSE) for the distance between the cover image \textbf{x} and the stego \textbf{y}. 
\begin{gather}
    dist(\textbf{x},\textbf{y})=\left(  {\sum}^{w}_{i=1}{\sum}^{h}_{j=1}(|n_{ij}|)\right)/w.h \, ,
\end{gather}One can notice that minimizing the loss using this distance, forces Agent-Alice to output only \textbf{zeros} over the map of modifications. Agent-Bob is then no longer capable of retrieving the secret message. To reduce this constraint, we introduce a constant $\beta$ in Agent-Alice's loss. This $\beta$ value is related to the change rate notion. So the loss becomes:
\begin{gather}
    \mathcal{L}_{Alice} = \lambda_A \cdot (dist(\textbf{x,y})-\beta)+ \lambda_B  \cdot\mathcal{L}_{Bob}-\lambda_E \cdot\mathcal{L}_{Eve},
    \label{eq:second archi}
\end{gather} where $\lambda_A, \lambda_B, \lambda_E\mathrm{\in } [0,1] $. Note that $\beta$ controls the discretion of the embedding network, i.e. how many pixels Agent-Alice is allowed to alter from the cover image \textbf{x}.  

\subsection{Third-Architecture (source separation)}
\begin{figure*}[htb]% height=4in,
\includegraphics[height=4in,width=7in, keepaspectratio=False]{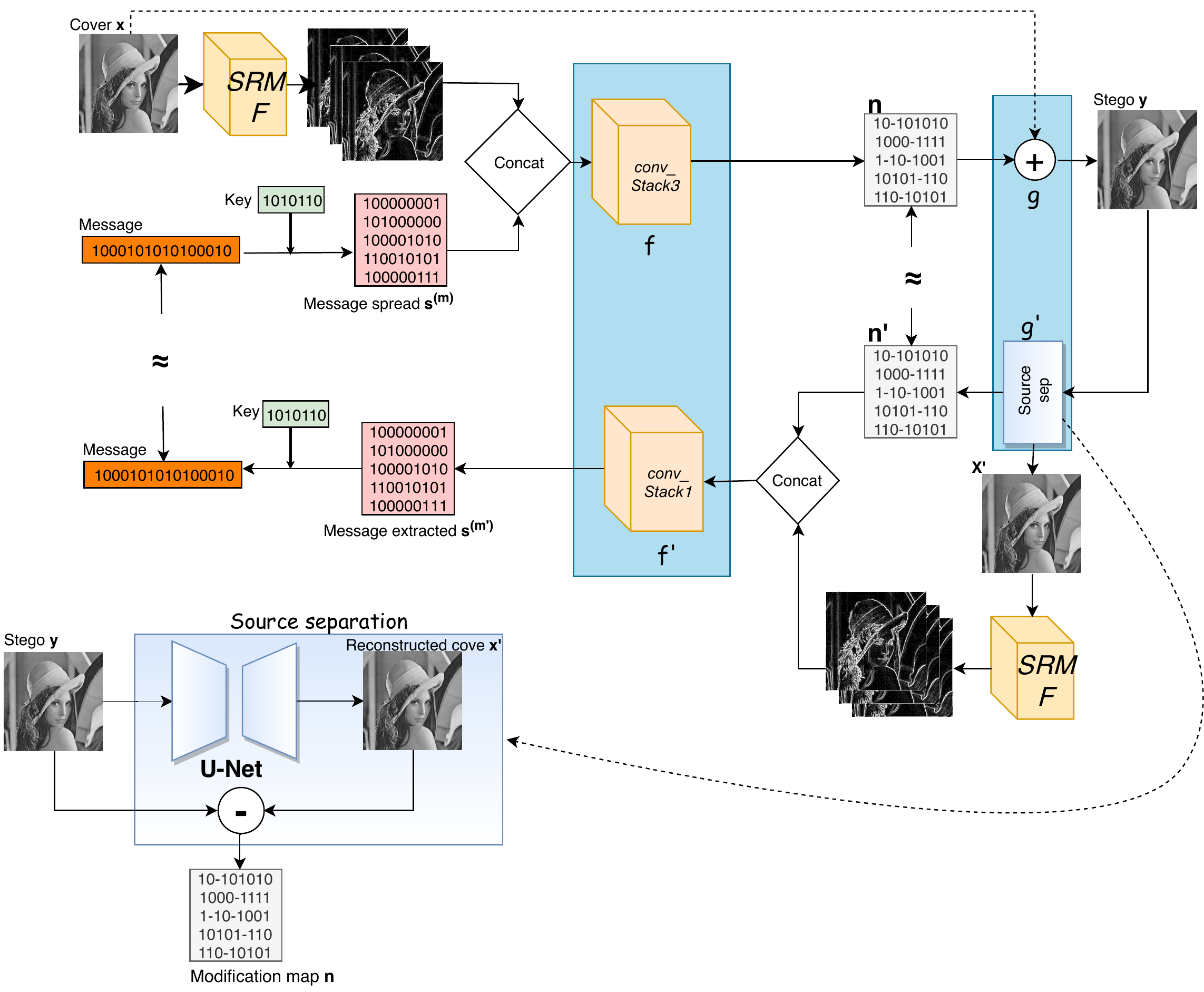}
\caption{Agent-Alice and Agent-Bob with the third architecture.}
\label{fig:third architecture}
\end{figure*} 

The architecture shown in Fig.~\ref{fig:third architecture} is proposed as an improvement to the second architecture. The embedding part of the second architecture was changed to make as few adjustments as possible. The extracting part, on the other hand, remains the same from the first architecture. Nevertheless, constraining the amplitude of modifications directly impacts the extracting part. When we limit the number of pixels that can be modified, more errors occur during message extraction. In other words, altering fewer pixels means less detectability, but more errors during \textit{message-extraction}, while changing more pixels means fewer errors when retrieving the message, but more detectability. How can Agent-Bob extract the secret message correctly when Agent-Alice carries out the minimal required modification?

In embedding algorithms such as S-UNIWARD \cite{Holub2014_S-UNIWARD}, WOW \cite{Holub2012_WOW}, etc, the message coding requires, in practice, the use of a Syndrome-Trellis-Codes STC \cite{STC}. The extractor (Bob) has access to the parity-check matrix \textbf{h}$\mathrm{\in }\mathrm{\{0,\textit{1}\}}^{w\times h}$ used by Alice during the embedding process. This matrix \textbf{h} is then shared between Alice and Bob, so Bob can easily retrieve the message from \textbf{y} by calculating the matrix product as:
\centerline{\textbf{m}= \textbf{h} $\cdot$ lsb(\textbf{y}),}
with lsb (.) the function extracting the LSB plane.

In our proposed method, no parity-check matrix is shared between Agent-Alice and Agent-Bob, nor any related information. If Agent-Bob has to mimic an STC to perform the message extraction in the \textit{3-player game}, it should learn a matrix in conjunction with Agent-Alice, in order to retrieve the message correctly. Without any topological or mathematical construction, this can be difficult, especially when the system used has many loss terms to minimize (Nash equilibrium issue).

A solution for this problem could be to inject a coding/decoding block inside Agent-Alice and Agent-Bob. Nevertheless, this is not an easy task, and before trying this solution, we preferred exploring the impact of increasing the link between the two agents. The integration of a coding and decoding block is postponed for future work.

We continue in the direction of increasing the link between Agent-Alice and Agent-Bob by forcing their topology to be more anti-symmetric. Referring to Fig.~\ref{fig:third architecture}, we draw two blue rectangles to show where this anti-symmetry has been injected. On one of the rectangles, we note {\bf \it{g}}, the {\it point-wise summation} block, and {\bf \it{g'}}, the {\it sources separation} block. {\bf \it{g}} and {\bf \it{g'}} can be seen as inverse functions such that: 
\begin{eqnarray*}
g : \{-1, 0, +1\}^{w\times h} \times \mathrm{\mathbb{R}}^{w\times h} \rightarrow \mathrm{\mathbb{R}}^{w\times h}\\
g({\mathbf n}, {\mathbf x})   = {\mathbf n} + {\mathbf x} = {\mathbf y}\\
g' : \mathrm{\mathbb{R}}^{w\times h} \rightarrow \{-1, 0, +1\}^{w\times h} \times \mathrm{\mathbb{R}}^{w\times h}\\
g'({\mathbf y}) = ({\mathbf n'} , {\mathbf x'})
\end{eqnarray*}
For a given cover, {\bf x}, and a stego noise, {\bf n}, $g'( (g({\bf n}, {\bf x}) )$ gives an estimated cover, {\bf x'}, and a estimated stego noise, {\bf n'}. During learning, we then minimize the distance between {\bf x} and {\bf x'} (which is equivalent to minimizing the distance between {\bf n} and {\bf n'}); See Eq.~\ref{eq:lbob_cover}.

The {\it sources separation} block act as a denoiser. The well-known U-Net \cite{U-Net} has been used and integrated into Agent-Bob's architecture (see Fig.~\ref{fig:third architecture}). The goal of the U-Net is to reconstruct the cover image \textbf{x} from the stego \textbf{y} image. The modification map \textbf{n} is resulting from the subtraction of the reconstructed cover \textbf{x'} from the stego image \textbf{y}. 

Referring to Fig.~\ref{fig:third architecture}, on the other rectangle, we note {\bf \it{f}}, the ${\it conv\_Stack3}$ block, and {\bf \it{f'}}, the ${\it conv\_Stack1}$ blocks. {\bf \it{f}} and {\bf \it{f'}} can be seen as inverse functions from the point view of the spread message ${\mathbf s^{(m)}}$ and the modification map ${\bf n}$ :
\begin{eqnarray*}    
f : \mathrm{\mathbb{R}}^{w\times h\times30} \times \mathrm{\{0,1\}}{}^{w\times h}\rightarrow \{-1, 0, +1\}^{w\times h}\\
f_{\mathbf{r}}(\mathbf{s^{(m)}}) = \mathbf{n}\\
f' : \mathrm{\mathbb{R}}^{w\times h\times30} \times \{-1, 0, +1\}^{w\times h}\rightarrow \mathrm{\{0,1\}}{}^{w\times h}\\
f'_{{\mathbf r'}}({\mathbf n'}) = {\mathbf s'^{(m')}}
\end{eqnarray*}
with {\bf r} and {\bf r'} the 30 residual images obtained by $\textit{SRM-F}$ filtering. 

For a given spread message ${\mathbf s^{(m)}}$, and residual images {\bf r} and {\bf r'}, $f'_{{\mathbf r'}}( (f_{\mathbf{r}}({\bf s^{(m)}}) )$ gives an estimated spread message ${\mathbf s'^{(m')}}$. During the learning, we then minimize the distance between {\bf m} and {\bf m'} (which isthe equivalent to minimizing the distance between ${\mathbf s^{(m)}}$ and ${\mathbf s'^{(m')}}$; see Eq.~\ref{eq:Bob_mask}); See Eq.~\ref{eq:lbob_message}. Note that \textit{conv\_Stack1} and \textit{conv\_Stack3} are the same as in the second architecture.

Therefore Agent-Bob's loss is composed of two loss terms:
\begin{itemize}
\item the cover reconstruction loss:
\begin{equation}
\mathcal{L}_{cover\_recons} = MSE(\textbf{x'},\textbf{x})\,
\label{eq:lbob_cover}
\end{equation}
\item the message extraction loss:
\begin{equation}
\mathcal{L}_{message\_extract}= MSE(\textbf{m},\textbf{m'})\,
\label{eq:lbob_message}
\end{equation}
\end{itemize}
Agent-Bob loss is given as:
\begin{equation}
\mathcal{L}_{Bob}=1/2 (\mathcal{L}_{cover\_recons}+ \mathcal{L}_{message\_extract})\
\end{equation}

\section{Experiments}
\label{sec:experiments}

\subsection{Dataset and software platform}

In this paper, the experiments are carried out on two image sources. The first is the well-known database BOSSBase 1.01 \cite{Bas2011-BOSS} which contains 10,000 8-bit grayscale 512$\times$512 pixels sized images. This database was created for steganalysis purposes in 2011. Images from this database offer different texture characteristics, which explains why it is widely used in steganalysis. BOWS2 \cite{BOWS2008} is our second database. It was created for a watermarking contest and consists of 10,000 8-bit grayscale 512$\times$512 pixels sized images.

Due to our GPU computing platform, time limitation and nature of \textit{3-player game}, which takes much time to train due to two-phase learning. We conduct all the experiments on images of 256$\times$256 pixels. To this end, we resampled all images from the two databases from 512$\times$512 pixels to 256$\times$256 pixels, using the imresize() Matlab function with default parameters.\\
We implemented the proposed architectures presented in section \ref{sec:proposed_Architectures} using TensorFlow V 1.6. As for comparison, we use S-UNIWARD \cite{Holub2014_S-UNIWARD}, and WOW \cite{Holub2012_WOW}, two well-known content-adaptive methods for spatial domain embedding. All our experiments were conducted on the NVIDIA Titan X GPU platform.%\footnote{Source code will be made available.}.

\subsection{Training, Validation, Test}
\label{Training_section}
We start the training phase by preparing a pre-trained model of Agent-Eve. For this, we use the S-Uniward algorithm to generate 10,000 stegos from the BOSSBase. 10,000 cover/stego pairs are then obtained. We use 4,000 pairs to train a Yedroudj-Net network. After several epochs of training, we obtain a learned model of Yedroudj-Net, which we transfer to Agent-Eve; Agent-Eve does not learn from scratch. 

Once Agent-Eve is pre-trained, we start the learning phase of the 3-players (Agent-Alice, Agent-Bob and Agent-Eve).  We use BOSSBase as the image source. The messages are generated using a PRNG \footnote{PRNG: Pseudo-Random Number Generator. The PRNG generates a binary vector {\bf m} whose values 0 and 1 are uniformly distributed.} with a chosen payload. We use Adam, an adaptive optimizer, to train our system for all three networks where the mini-batch size is set to 4. The values of  $\lambda_A$, $\lambda_B$, $\lambda_E$ are set to 0.2, 0.4, and 0,4 respectively. These values are chosen empirically. However, they may not be optimum.
\begin{figure*}[htb]
\centering
\includegraphics{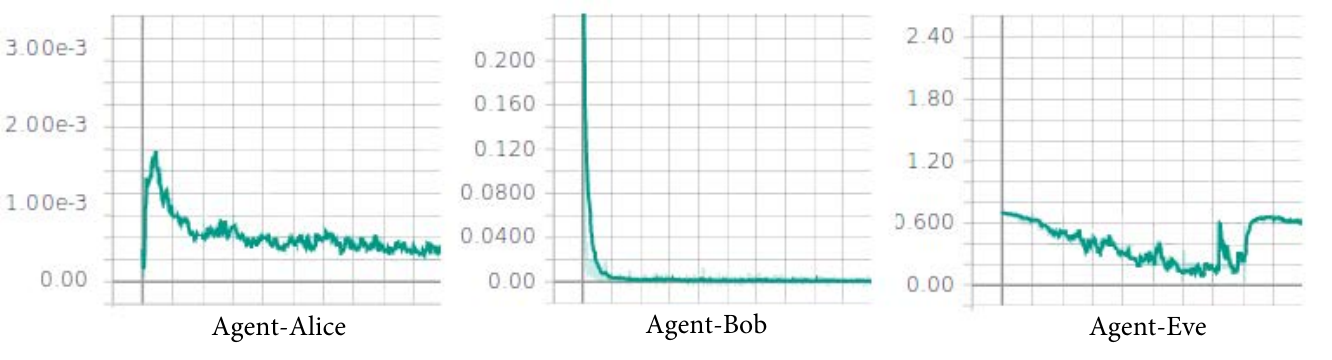}
\caption{Generic loss evolution of Agent-Alice, Agent-Bob and Agent-Eve after 300000 iterations.}
\label{fig:loss-curves}
\end{figure*}

During training, we set the maximum number of iterations \textit{max-iter} to 1 million iterations. The training is alternated between Agent-Alice and Agent-Bob from one side, and Agent-Eve from the other side. We set {\it it1} to 50 iterations, and {\it it2} to 1 iteration (see Algorithm.\ref{Algo:3players}, so Agent-Alice and Agent-Bob are trained for 50 iterations, compared to 1 iteration for Agent-Eve).

When the losses of the three networks appear to be stable, we freeze their training, and we integrate the discretization module into Agent-Alice (see Fig.~\ref{fig:The overall architecture}). Then, we resume the training for several iterations (more or less 50,000 iterations), so the system learns how to generate stego images with discrete pixel values.

We should note that activating the discretization module at the beginning of the training phase prevents the system from converging (as the round function is not differentiable). So, We let the system converge towards a solution, then we force it to work on images with discrete pixel values. Once our system is well trained (loss curves are stable as shown in Fig.~\ref{fig:loss-curves}), we stop the training phase. 

To evaluate the performance of our method, we measure the \textit{security} and the \textit{transmission errors}. The security is measured with the probability of error (Pe) of a given steganalyzer, where Pe is computed as the average of the false alarm rate and the missed detection rate. For the transmission error, we use the Bit Error Rate (BER)\footnote{Note that the BER is not useful in a standard steganographic scenario since no channel error should be considered during the transmission of the stego image. Unfortunately, none of the architecture of existing literature for the \textit{3-player game} are able to embed a message that could be retrieved without errors}. 

The reader should understand that Agent-Alice is embedding a {\it secret message}, noted {\bf m}, which is a binary vector. You should consider that this vector {\bf m} is the resulting of, first, the encryption of a {\it clear message}, with a classical cryptographic algorithm, and second, its encoding by an error-correcting code\footnote{In many embedding schemes, the encoding by an error-correcting code is optional.}. Making this assumption is natural, because it ensures that there are no security leaks in the secret message, {\bf m}, and that there is an equiprobability of 0 and 1 in the secret message.

So, for the evaluation of the {\it transmission error}, we use the Bit Error Rate (BER) such that BER = $dist({\bf m}, {\bf m'})/m$, with $dist(.)$, the Hamming distance. In this paper, the BER is thus the ratio between the number of bit errors in the {\it secret} binary message, {\bf m'}, {\it extracted} by Agent-Bob, on the number of bits from the {\it secret} binary message, {\bf m}, {\it embedded} by Agent-Alice. SO, Agent-Bob is  extracting a secret {\bf coded}-message, {\bf m'}, whith a BER which is possibly not null, but the secret {\bf decoded}-message will have a null BER, because Alice sends only stego images to Bob, whose {\bf decoded}-message is free from errors. 
 Note that we will not integrate the error-correcting code block, but we will discuss it. Indeed, this paper is primarily on the 3-player concept definition and the proposition of three architectures. The reader should consider that {\bf m'} is an encrypted and coded binary vector. All of this also implies that the ``payload rate'' (i.e. the embedding rate), is equals to $m/(w\times h)$, and the ``real payload rate'', is equal to the size of the {\bf decoded}-message divided by $w\times h$.

Therefore for the testing phase, Agent-Alice starts generating 10,000 Stego images from BOWS2, where a random message is embedded in a cover image using a random key. The generated images are used to evaluate the accuracy of Bob's message retrieval on the one hand and the security of our steganographic system against the steganalyzer Yedroudj-Net on the other. Please note that the steganalysis with Yedroudj-Net (learning and testing) is achieved on BOWS2Base, as is the training of the 3-players (Agent-Alice, Agent-Bob, and Agent-Eve) it is carried out using the different and disjointed BOSSBase. The results are presented in the following subsection. 

\subsection{Results findings and discussion }

\subsubsection{Extraction and security comparison of the three architectures}
\noindent

\textbf{First architecture:}

In Table \ref{tab:First_arch_results}, we report the Bit Error Rate (BER) and the Probability of error (Pe) obtained using the first architecture. These tests are carried out on BOWS2 database using different payloads 0.2 bpp, 0.4 bpp and 1 bpp. The steganalyzer used is Yedroudj-Net.
Regardless of the payload, Bob, who uses the Agent-Bob can correctly recover the secret message with a BER equal to zero. We can observe that this architecture is not at all secure since the detection accuracy obtained using Yedroudj-Net is between 96\% and almost 98\%.

The first architecture offers good embedding capacity, and we can perfectly recover the secret message even when the payload is significant, although the low security given by this architecture does not make it interesting for steganography purposes.

Note that this architecture is the closest form of architecture to GSIVAT or HiDDeN since the main principle is to spread the message in the spirit of a spread spectrum watermarking approach.
This first architecture, GSIVAT, and HiDDeN are definitively insecure approaches since the stego noise power is too strong. In the second architecture We investigated a way to constrain pixels modifications to -1 or +1.

\begin{table}
\centering
\caption{First architecture BER and Pe for different payloads.}
\renewcommand{\arraystretch}{1.0}
\scalebox{1.}{

\begin{tabular}{|c|c|c|}
\hline
\multicolumn{1}{|c|}{\textbf{Payload}} & \multicolumn{1}{c|}{\textbf{Bite Error Rate}} & \multicolumn{1}{c|}{\textbf{Probability of error}} \\ \hline
0.2                                    & {\color[HTML]{009901} 0}        & {\color[HTML]{CB0000} 3.7\%}     \\ \hline
0.4                                    & {\color[HTML]{009901} 0}        & {\color[HTML]{CB0000} 3.3\%}     \\ \hline
1                                      & {\color[HTML]{009901} 0}        & {\color[HTML]{CB0000} 2.5\%}     \\ \hline
\end{tabular}}

\label{tab:First_arch_results}
\end{table}

\vspace{1cm}
\textbf{Second architecture:}

\begin{figure}[htb]
\includegraphics*[width=3.4in, height=2.in, keepaspectratio=false]{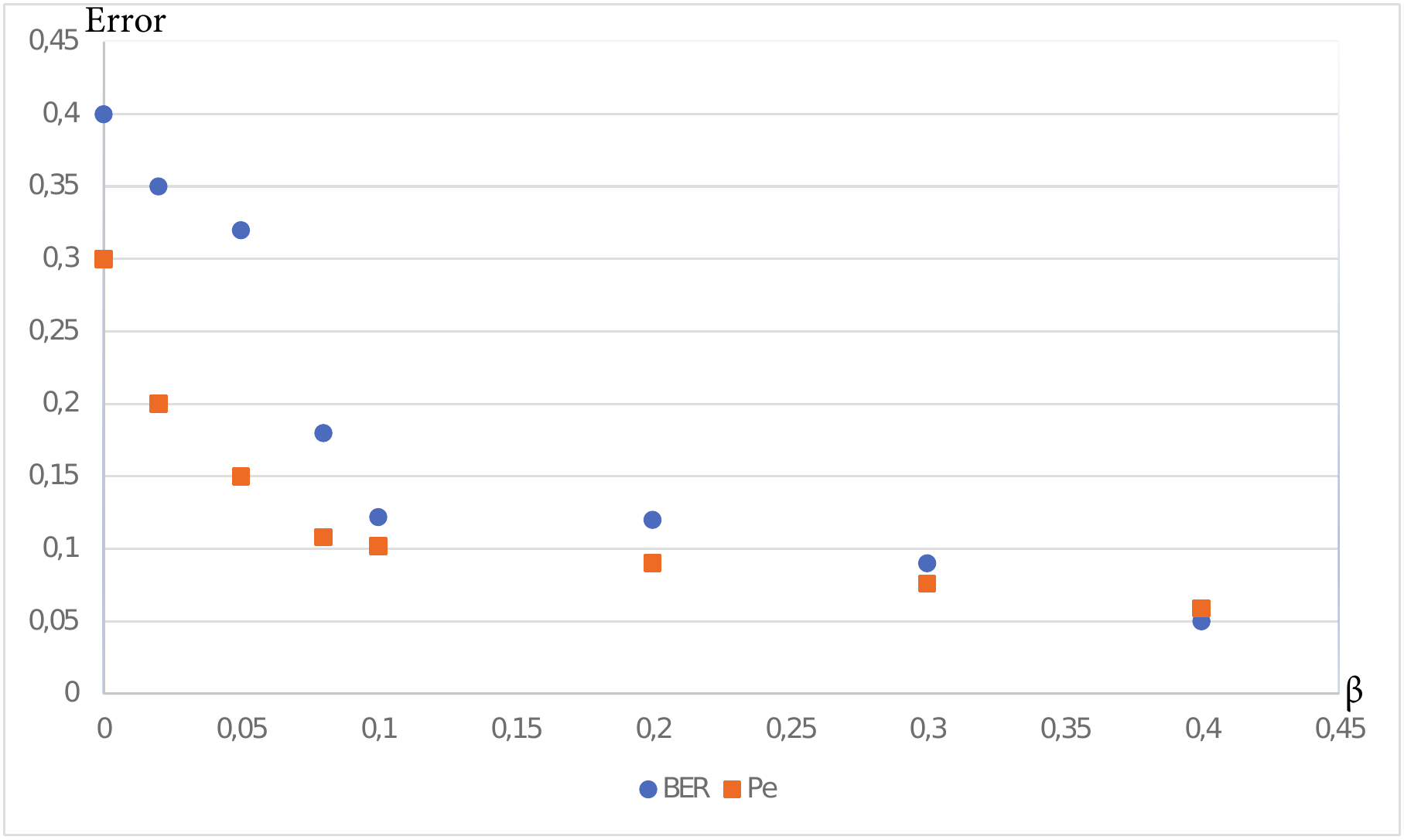}
\caption{BER and Pe for the second architecture for a payload size of 0,4 bpp in function of change rate $\beta$.}
\label{fig:Second-arch-results}
\end{figure}

In the second architecture, we manage to control the noise power introduced by Agent-Alice during the embedding process. To do this, we have conducted several experiments with different values of the change rate $\beta$ of the Eq.~\ref{eq:second archi}. 
In Fig.~\ref{fig:Second-arch-results}, we present the Bit Error Rate (BER) and the Probability of error (Pe) of the second architecture, when different $\beta$ values are used, for a fixed payload of 0.4 bpp. The used steganalyzer was Yedroudj-Net.
We can see that, for $\beta$ equal to 0.4, Eve, the steganalyzer (Yedroudj-net) gets a Probability of error (Pe) close to 6\% when Bob makes a 5\% Bit Error Rate (BER). When $\beta$ is set to 0.2, the security of our system improves with a Pe equal to 9\%, although the message extraction becomes more difficult for Bob with a 12\% Bit Error Rate. 

For a $\beta$ value equal to or less than 0.05, the generated images are more secure. Nevertheless, the BER increases considerably. The BER is 32\% and 40\%, and the Pe is 15\% and 30\% for $\beta$ equals 0,05 and 0 respectively. These BER values are very high and would be difficult to correct with a correcting code without strongly reducing the real payload size. The best $\beta$ value is those that in the same time, maximize the detection error value (Pe), and minimize the size of the coded message; the message has to be encoded such that a correct extraction is ensured. Since these considerations are a little bit of topic, we propose a rapid analysis by choosing a particular point where there is a sudden variation of BER (i.e. the inflection point). For this architecture and for a payload size 0.4 bpp, the value of $\beta$ corresponding to a sudden variation of BER is around $\beta = 0.1$. At $\beta = 0.1$, the BER=12\%. Using the [7,4,3] Hamming Error Correcting Code (ECC) ensures a correction to most 14\% BER, so we would, on average, correct all the errors. With this ECC the {\it real} payload size is 0.23 bpp, and we measure for this point a probability of Error of 10.2\%. It is clear that this architecture provides better results compared to the first one, where for a payload of 0.2 bpp the Pe is 3\%. Nevertheless, these results are still not convincing, and as explained before, the problem of the second architecture is the weak relationship between Agent-Alice and Agent-Bob. The third architecture aims to counter this weakness.

\begin{figure}[htb]
\includegraphics[width=3.4in, height=1.in, keepaspectratio=false]{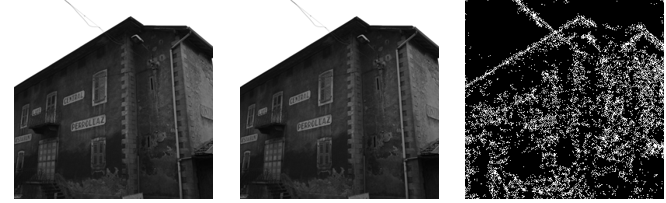}
\caption{(Left) The BOSSBase cover image. (Middle) the corresponding stego images with 0.4 bpp and $\beta =$ 0.1 using the second architecture.  (right) the modification maps between the cover image and the corresponding stego where black=0, white=+/- 1.
}
\label{fig:modif_map_archi2}
\end{figure}

In Fig.~\ref{fig:modif_map_archi2} we can observe that this architecture can learn to concentrate the embedding on textured regions, which are more difficult for a steganalyzer to detect. This architecture has the capacity to learn and find interesting zones to implement steganography.

\textbf{Third architecture: }
\begin{figure}[htb]
\includegraphics*[width=3.4in, height=2.in, keepaspectratio=false]{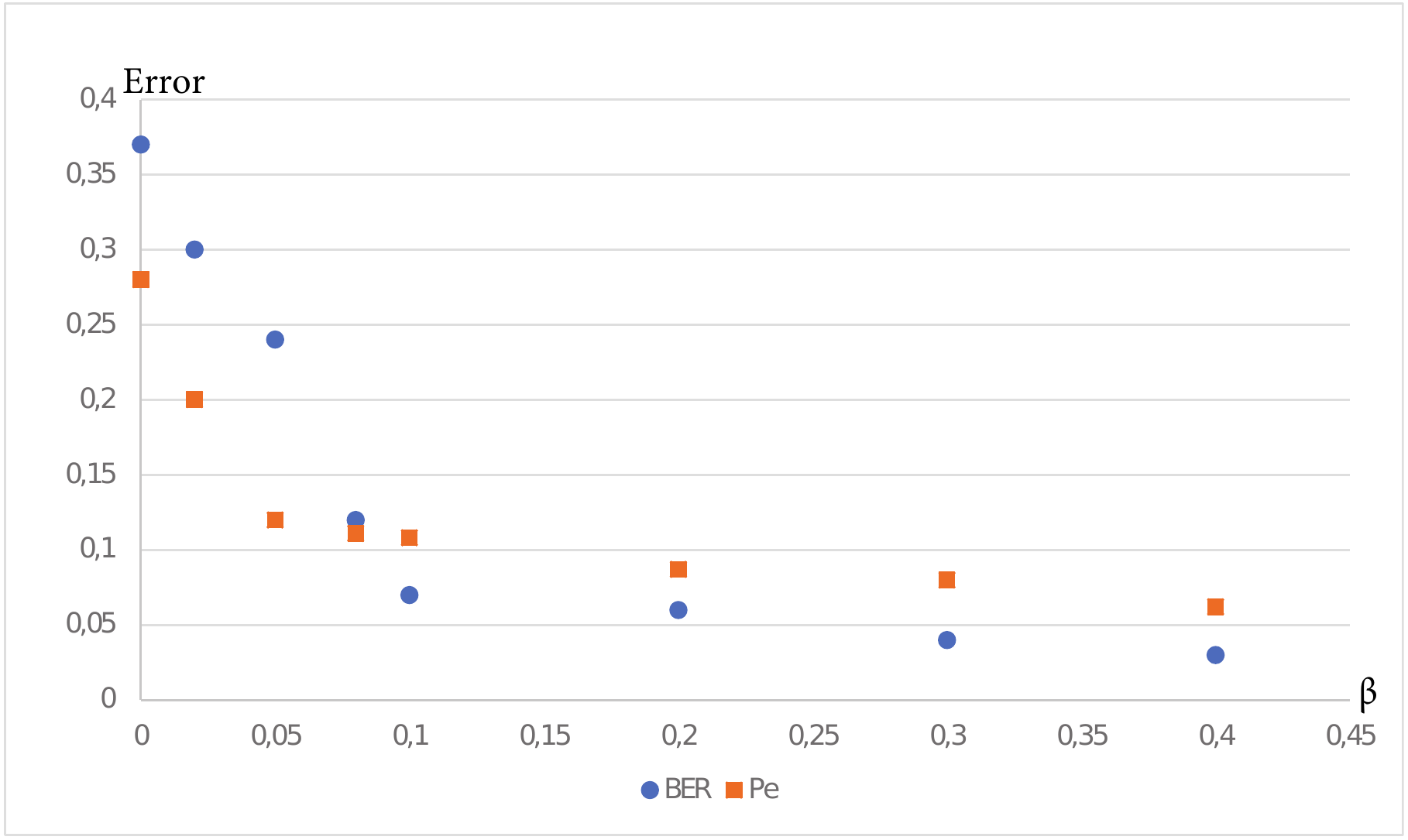}
\caption{BER and Pe for the third architecture for a payload size of 0.4 bpp in function of change rate $\beta$.}
\label{fig:third_arch_results}
\end{figure}

As previously mentioned, the third architecture has been proposed to improve the extraction part of the second architecture. However, this requires more time to converge, but the up side is that it offers a better message retrieval accuracy.
In Fig.~\ref{fig:third_arch_results}, we present the bit error rate (BER) and the probability of error (Pe) of the third architecture as a function of $\beta$ the change rate, for a fixed payload 0.4 bpp. 

Compared to the second architecture, the error of Bob (BER) is smaller regardless of the value of $\beta$, even though the detection accuracy remains almost the same for both architectures. The BER obtained with the third architecture and in comparison with the second architecture is 2\% lower when $\beta=0.4$, 6\% lower when we set $\beta= 0.2$, and 5\% lower when $\beta= 0.1$.
By observing these results, we see that Agent-Alice and Agent-Bob have nevertheless benefited from the third architecture. More joint learning between Agent-Alice and Agent-Bob seems to be the right research direction.

Looking at Fig.~\ref{fig:third_arch_results}, similarly to the second architecture, the inflection point is around $\beta =0.1$. At this point, the BER value is 0.06. The errors can be corrected with a Hamming code [15,11,3].

In this case the {\it real} payload size is $(0,4\cdot11)/{15}= 0.293 bpp    \approx0.3$ bpp. Using this payload we get a Pe of about 11\%. 
 
To provide a comparison, we run a steganalysis with Yedroudj-Net against two steganographic algorithms S-UNIWARD and WOW at a payload size of 0.3 bpp using the database BOWS2. 
The probability of error (Pe) is 27.3\% (resp. 22.4\%) for S-UNIWARD (resp. WOW). There is undoubtedly a security gap with the actual embedding schemes, but again, the objective of the paper is to define the {\it 3-player game} concept, and to analysis, its potential compared to other modern embedding schemes and steganalysis scheme. In this, the third architecture shows that there is a real potential, and this paper paves the way to many research possibility.

\begin{figure}[htb]
\includegraphics[width=3.4in, height=1.in, keepaspectratio=false]{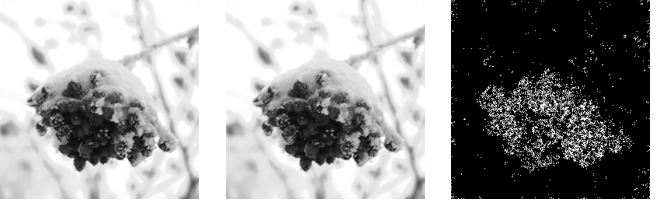}
\caption{(Left) The BOSSBase cover image. (Middle) The stego image with payload size 0.4 bpp and $\beta=0.1$ using the third architecture.  (right) the modification maps between the cover image and the corresponding stego where black=0, white=+/- 1.
}
\label{fig:modif_map_archi3}
\end{figure}

Fig.~\ref{fig:modif_map_archi3} shows the modifications zone using the third architecture. We can observe some adaptivity, as the embedding is more dense in textured zones.

The results show that this architecture can implement adaptive embedding. It also indicates that the link between Agent-Alice and Agent-Bob should be reinforced in the future to reduce the BER and eventually reduce the change rate and increase the security level. These results should not deter from the 3-player concept, even if the security level at the moment is lower compared to the traditional adaptive embedding algorithms such as S-UNIWARD and WOW. Indeed, the paper shows real improvements compared to GSIVAT and HiDDEeN.  

\subsubsection{Model's security analysis}
\noindent
To further investigate the security of the proposed model additional tests have been carried out. We list below the points we wanted to investigate with these tests, with a brief description for each of them:

%\begin{description}
\vspace{0.2cm}	
\textbf{Using a different steganalysis algorithm for the testing phase:}
    In previous experiments, we used the Yedroudj-Net (for playing the role of Agent-Eve) in order to train Agent-Alice and Agent-Bob, but also to evaluate Agent-Alice's security level at test time. To this end, and in order to determine whether the embedding algorithm, Agent-Alice, has a similar security level against other steganalysts, we use CovPool-Net \cite{Deng2019_Cov-Pool}, which is a recent state-of-the-art steganalyzer, that was published in July 2019. Thanks to its deep architecture design, this network achieves a good level of performance. The results are presented in Table \ref{tab:steganalyst_inf}.
    \begin{table}[htb]
\centering
\caption{Steganalysis error probability obtained with Yedroudj-Net and CovPool-Net on BOWS2 base. For data embedding, the second and third architectures are used at payload 0.4 bpp.}
\begin{tabular}{l|c|c|}
\cline{2-3}
 &
  \multicolumn{1}{l|}{Yedroudj-Net} &
  \multicolumn{1}{l|}{CovPool-Net} \\ \hline
\multicolumn{1}{|l|}{\begin{tabular}[c]{@{}l@{}}Third architecture / \\ real payload 0.3bpp\end{tabular}} &
  10.8 \% &
  11.3 \% \\ \hline
\multicolumn{1}{|l|}{\begin{tabular}[c]{@{}l@{}}Second architecture /\\  real payload 0.23bpp\end{tabular}} &
  10.2 \% &
  10.2 \% \\ \hline
\end{tabular}
\label{tab:steganalyst_inf}
\end{table}
    As one can observe, the error probability obtained with Yedroudj-Net and CovPool-Net on BOWS2, whatever the architecture, are similar. These results show that Agent-Alice is not specifically tuned to defeat Yedroudj-Net (Agent-Eve), otherwise, CovPool-Net would achieve a smaller error probability. Indeed, CovPool-Net is obtaining similar results compared to SRNet \cite{Deng2019_Cov-Pool}, and SRNet is slightly better than Yedroudj-Net \cite{Zhu-Net}; meaning that CovPool-Net is supposed to be slightly better than Yedroudj-Net. So, our embedding algorithm (Agent-Alice), seems sufficiently generic and does not show security leaks when steganalyzed with a steganalysis algorithm different from Yedroudj-Net.

\vspace{0.2cm}	
{\bf Using another database to evaluate the security level:} The purpose of this test is to study the influence of the database on the security level of the embedding algorithm, Agent-Alice. That is to say, during the test phase, instead of using a database (BOWS2 base) that is statistically similar to the one used during the training phase (BOSS base), we used a database whose statistics are different. For this, we used the ALASKA-10K base \cite{Yedroudj_Pixels_off}. Please note that this database is more secure than BOSS and BOWS2 i.e. more difficult to steganalyze.
    
    \begin{table}[htb]
    \centering
    \caption{The BER and steganalysis error probability of Yedroudj-Net on BOWS2 and ALASKA-10K bases, with an embedding with Agent-Alice obtained using the third architecture on BOSSBase with a payload 0.4 bpp.
}
    \label{tab:my-table}
    \begin{tabular}{l|l|l|}
    \cline{2-3}
                                 &  Probability of error &  Bite Error Rat \\ \hline
    \multicolumn{1}{|l|}{BOWS2}      & 10.8 \%                   & 6 \%                 \\ \hline
    \multicolumn{1}{|l|}{ALASKA-10K} & 12.1 \%                   & 13 \%                \\ \hline

    \end{tabular}
    \label{tab:base_inf}
    \end{table}

    Table \ref{tab:base_inf} shows that for an embedding with the same modification rate, ALASKA-10K is a more secure database than BOWS2, which is consistent with previous studies \cite{Yedroudj_Pixels_off}. Indeed, using this database results in a 1\% increase in Pe compared to BOWS2. However, this database produces more decoding errors. Indeed, the BER is 13\% for Alaska-10K versus 6\% for BOWS2. Thus, the correction ability of a Hamming code [15,11,3] is no more sufficient. We can, for example, use a Hamming code [7,4,3]. In thus case, the real payload is reduced to 0.23 bpp on ALASKA-10K, compared to 0.3bpp on BOWS2.

\vspace{0.2cm}	
{\bf Testing different payloads:}
    The objective of this test is to analyze the efficiency of Agent-Alice, when the embedding payload size is not the ``designed payload size''. Indeed, Agent-Alice is trained at a given payload, and it is conceptually designed to be capable of supporting embedding at different payload sizes during its use in the test phase. So, the objective of this test is to study if Agent-Alice's embedding algorithm obtains relatively good results when embedding at a higher or lower payload size. To do this, we trained Agent-Alice on a payload of 0.4 bpp, and then, thanks to the scalability property of our model, we generated stego images with different payload sizes (0.14 bpp, 1bpp).
    
    \begin{table}[htb]
    \centering
\caption{Steganalysis error probability of Yedroudj-Net on BOWS2 base. For data embedding the third architecture is used at different payloads.}
\label{tab:Payload_inf}
\begin{tabular}{|l|l|l|}
\hline
Payload size (bpp) & Real   payload size (bpp) & Pe     \\ \hline
0.14                 & 0.1                       & 11.1\% \\ \hline
0.4                  & 0.3                       & 10.8\% \\ \hline
1                    & 0.8                       & NaN    \\ \hline
\end{tabular}
\end{table}

    Table \ref{tab:Payload_inf} illustrates that our method achieves the same level of security for both payloads (0.14 and 0.4 bpp). However, when the payload is equal to 1, Agent-Bob is no longer able to extract the message. This last result suggests that the quantity of modification is insufficient. So, the change rate $\beta$ (Eq. \ref{eq:second archi}) has to be modified, and the learning of Agent-Alice has to be re-done for a higher payload. When embedding with a lower payload, the network can successfully embed and extract the message, but the security level remains the same as the one obtained for the payload size used during the training phase. For better security performances, the change rate $\beta$ (Eq. \ref{eq:second archi}) should be changed, and at worst, we should relaunch training with the target payload size. Another solution, in order to ensure ``payload scalability'' could be  ``transfer learning'', or
     to run learning with variable payload sizes. This is postponed to future studies.
   
\vspace{0.2cm}	
{\bf Embedding with different secret keys:} A major weakness of previous models lies in the use of a single secret key for data embedding, even though it is highly discouraged to do so in steganography, as leads to very high detectability \cite{Pibre2016}. To avoid such a limitation, we propose a dedicated solution that integrates the use of a different key for each data embedding process (stego generation).
    
    To prove that our model handles the secret key in the way it is intended to, we run the embedding process twice by using the same image and the same message, but with two different secret keys.
    
    \begin{figure}[htb]
    \hspace{0.3in}
\includegraphics[width=3.2in, height=1.in, keepaspectratio=false]{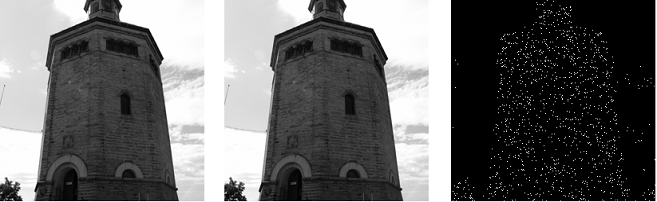}
\caption {(Left) The stego image generated with secret key 1 . (Middle) The stego images generated with secret key 2. (right) the
difference between the two stego images where black=0, white=+/- 1.
}
\label{fig:key_influence}
\end{figure}

    Fig. \ref{fig:key_influence} demonstrates that the model was able to generate a completely new stego image by only changing the secret key. This proves that the proposed method makes correct use of the shared secret key.

%\end{description}

\section{Conclusion and perspectives}
\label{sec:Conclusion}

In this paper, we first recalled the four different GAN families used in steganography. Then we presented the \textit{3-player game approach}, and defined the general concept and how to correctly use it for steganography. Three architectures based on the \textit{3-player game approach} have been proposed. The first architecture fixed the flaws made in GSIVAT and HiDDEeN. Nevertheless, this first architecture behaves similarly to GSIVAT and HiDDEeN and is not adapted for steganography purposes due to its extreme detectability. With the second architecture, we suggested a new way to embed a message in the cover image. Instead of directly modifying the cover image, which implies a significant noise power signal, we proposed to generate a modification map with values belonging to \{-1,0,1\}. The stego is then generated by adding the modification map to the cover. This architecture is much more secure than the first one, but it generates more errors during message extraction. Finally, the third architecture imposes a more joint learning approach between Agent-Alice and Agent-Bob in order to reduce the errors during message extraction. This third architecture takes more time to converge, but achieves better results. 

The third architecture, with a {\it real} payload size of 0,3 bpp, can perfectly retrieve an embedded message (with the help of error-correcting code), for a security level Pe = 10,8\%. 
Visually speaking, see Fig. [\ref{fig:modif_map_archi2},\ref{fig:modif_map_archi3}] demonstrate that the proposed model was able to learn how to focus the embedding on textured regions, which are known to be more difficult to detect by a steganalyst. Thus, a better level of security is achieved.

Even if the obtained results do not surpass current state-of-the-art embedding algorithms, these experimental results verify the promises of such a method. The major contribution of this paper is really to propose a formalization of the {\it 3-player game} concept, and an end-to-end method using neural networks that can learn to simulate algorithms using human-based rules (steganography and steganalysis).

We expect this work to lead to fruitful avenues for further research.  In future work, we can study the possibility of more synchronizing between Agent-Alice and Agent-Bob, and therefor improving general performances. We can also try using a more subtle loss function. This can help the networks to converge to a better solution. Furthermore, finding a theoretical way to compute the change rate $\beta$ can help to accelerate the learning process. The values of  $\lambda_A$, $\lambda_B$, $\lambda_E$ could also be more extensively studied. 

\section*{Acknowledgments}

The authors would like to thank the University of Montpellier (LIRMM) and the University of N\^imes. We extend our gratitude to the French Direction G\'en\'erale de l'Armement (DGA) for its support through the Alaska project ANR (ANR-18-ASTR-0009). We would also like to thank the Algerian Ministry of Higher Education / Scientific Research, for its scholarship support.

\bibliographystyle{IEEEtran}
\bibliography{refs}

\end{document}